\documentclass{aa}
\usepackage{graphicx}
\usepackage{txfonts}
\def\approxgt{\mathrel{\hbox{\rlap{\lower.55ex \hbox {$\sim$}}
        \kern-.3em \raise.4ex \hbox{$>$}}}}
\def\approxlt{\mathrel{\hbox{\rlap{\lower.55ex \hbox {$\sim$}}
        \kern-.3em \raise.4ex \hbox{$<$}}}}

\begin{document}
   \title{The XMM-Newton/INTEGRAL monitoring campaign of IGR~J16318-4848}

   \author{A.Ibarra
          \inst{1},
          G.Matt
          \inst{2},
          M.Guainazzi
          \inst{3},
          E.Kuulkers
          \inst{4},
          E.Jim\'enez-Bail\'on,
          \inst{2}          
	  J. Rodriguez,
	  \inst{5}
	  F. Nicastro,
	  \inst{6}
	  R. Walter
	  \inst{7}
          }

   \offprints{A. Ibarra}

   \institute{$^1$XMM-Newton Science Operations Center, European Space Astronomy Center, INSA, Apartado
              50727, E-28080 Madrid, Spain \\
              \email{Aitor.Ibarra@sciops.esa.int} \\
              $^2$Universit\'a degli Studi ``Roma Tre'', Via della Vasca Navale 84, I-0046, Roma, Italy \\
	      $^3$XMM-Newton Science Operations Center, European Space Astronomy Center, ESA, Apartado
              50727, E-28080 Madrid, Spain \\
              $^4$Integral Science Operations Center, European Space Astronomy Center, ESA, Apartado
              50727, E-28080 Madrid, Spain \\
	      $^5$AIM - Unit\`e Mixte de Recherche CEA - CNRS - Universit\`e Paris VII - UMR 7158, 
	      CEA Saclay, Service d Astrophysique, F-91191 Gif sur Yvette, France\\
	      $^6$ Harvard-Smithsonian Center for Astrophysics, Cambridge, MA, 01238, USA\\
	      $^7$Integral Science Data Centre, Chemin d`Ecogia 16, CH-1290 Versoix, Switzerland
              }

   \date{Received ; accepted }

   \abstract {IGR~J16318-4848 is the prototype and one of the more
   extreme examples of the new class of highly obscured Galactic X-ray
   sources discovered by INTEGRAL. A monitoring campaign on this
   source has been carried out by XMM-Newton and INTEGRAL, consisting
   in three simultaneous observations performed in February, March and
   August 2004.}  {The long-term variability of the Compton-thick
   absorption and emission line complexes will be used to probe the
   properties of the circumstellar matter.} {A detailed timing and
   spectral analysis of the three observations is performed, along
   with the reanalysis of the XMM-Newton observation performed in
   February 2003. The results are compared with predictions from
   numerical radiative transfer simulations to derive the parameters
   of the circumstellar matter.} {Despite the large flux dynamic range
   observed (almost a factor 3 between observations performed a few
   months apart), the source remained bright (suggesting it is a
   persistent source) and Compton-thick ($N_{H} >1.2\times10^{24}$
   cm$^{-2}$). Large Equivalent Width (EW) emission lines from Fe
   K$_{\alpha}$, Fe K$_{\beta}$ and Ni K$_{\alpha}$ were present in
   all spectra.  The addition of a Fe K$_{\alpha}$ Compton
   Shoulder improves the fits, especially in the 2004
   observations. Sporadic occurrences of rapid X-ray flux risings were
   observed in three of the four observations. The Fe K$_{\alpha}$
   light curve followed the continuum almost instantaneously,
   suggesting that the emission lines are produced by illumination of
   small-scale optically-thick matter around the high-energy continuum
   source. Using the iron line EW and Compton Shoulder as diagnostic
   of the geometry of the matter, we suggest that the obscuring matter
   is in a flattened configuration seen almost edge--on.}{}


\authorrunning{Ibarra et al.}

\titlerunning{Monitoring of IGR~J16318-4848 with XMM-Newton}

   \maketitle
%

\section{Introduction}\label{Int}

IGR~J16318-4848 was discovered by INTEGRAL on January 29${^{\rm th}}$
2003, by the IBIS/ISGRI instrument, during a routine Galactic Plane
Scan (Courvoisier et al. 2003). Re-analysis of archival ASCA data
revealed at that position a highly absorbed
($N_{H} \sim 10^{24}~$cm$^{-2}$) source (Murakami et al. 2003, Revnivtsev
et al. 2003).

The discovery of this source triggered a XMM-Newton Target of
Opportunity observation on February 10, 2003, and the subsequent
analysis of the XMM-Newton data, in the 5-15 keV energy band,
confirmed the presence of a heavily absorbed source, $N_{H} \sim
2\times10^{24}$cm$^{-2}$, with strong emission lines (Schartel et al
2003). 

This complex emission could be resolved in three components, with
centroid energies of $6.410 \pm 0.003$ keV, $7.09 \pm 0.02$ keV and
$7.47 \pm 0.02$ keV, that correspond to Fe K$_{\alpha}$, Fe
K$_{\beta}$, and Ni K$_{\alpha}$ fluorescent emission lines
respectively (de Plaa et al. 2003).  These preliminary results also
showed that, during the XMM-Newton observation, IGR~J16318-4848 was
obscured by a Compton-thick absorber, $N_{H}=(1.66
\pm 0.16)\times10^{24}$ cm$^{-2}$. A detailed analysis of this
observation can be found in Matt \& Guainazzi (2003) and Walter et
al. (2003).

Further studies at different wavelengths have also been carried out on
the source. Filliatre \& Chaty (2004) found the optical counterpart of
IGR J16318-4848. Their studies suggest that the source is a High Mass
X-ray Binary (HMXB) at a distance between $~1$ and $~6$ kpc, the mass
donor being an early-type star, probably a $sgB[e]$ star, surrounded
by a dense and absorbing circumstellar material.  The $sgB[e]$ stars
are massive, evolved, high-luminosity stars undergoing mass loss in a
two-component wind (see, e.g., Hynes et al. 2002).

 The prominent, large equivalent width (EW) of the Fe
K$_{\alpha}$, Fe K$_{\beta}$, and Ni K$_{\alpha}$ fluorescent emission
lines, observed in the XMM-Newton spectrum of IGR~J16318-4848, makes
this object one of the most extreme examples of the new class of
highly absorbed X-ray binaries discovered by INTEGRAL [see, e.g,
Kuulkers, 2005 for a review].

A XMM-Newton observation campaign was carried out during 2004. At the
same time, INTEGRAL was also observing the source, collecting as well,
data at energies larger than 10 keV. Due to its prominent, large
EW Fe and Ni emission lines, IGR J16318-4848 is an almost unique
laboratory to study the physical condition and geometrical
distribution of matter accreting on the compact object in a binary
system. Thanks to this monitoring campaign we are able to compare the
results of the radiation transfer simulations by Matt et al. (1999)
and Matt (2002) to the spectra obtained through the coordinated
observations, in order to determine the time evolution of the
following quantities characterizing the accretion flow: column density
along the line of sight, covering fraction of the line-emitting
matter, some constraints on the geometry, and ionization state of the
matter.

This paper is organized as follows: in section ~\ref{data} information
about the observations and the data reduction are given,  while in
sections ~\ref{results} and ~\ref{discussion} the results are presented
and discussed.

\begin{table}
\caption{XMM-Newton monitoring campaign observations on IGR J16318-4848. 
The table shows EPIC-pn exposures times and source extraction region
around the object for each observation.}
\centering
\begin{tabular}{cccc}
\\
\hline\hline
\multicolumn{2}{c}
{\bf Start Time - End Time (UT)}  & {\bf T$_{\rm exp}$}  & {\bf{Region}} \\ \hline \hline
2003-02-10 16:58 & 2003-02-10 23:55 &  23000 s & 30$^{''}$ \\
2004-02-18 02:43 & 2004-02-18 08:16 &  17872 s & 30$^{''}$ \\
2004-03-20 22:13 & 2004-03-21 04:47 &  20213 s & 22.5$^{''}$ \\
2004-08-20 04:30 & 2004-08-20 10:04 &  18936 s & 30$^{''}$ \\
\hline
\end{tabular}
\label{tab1}
\end{table}


\begin{figure*}
\hbox{
\hspace{-0.1cm}
\includegraphics[width=1.0\textwidth, angle=0]{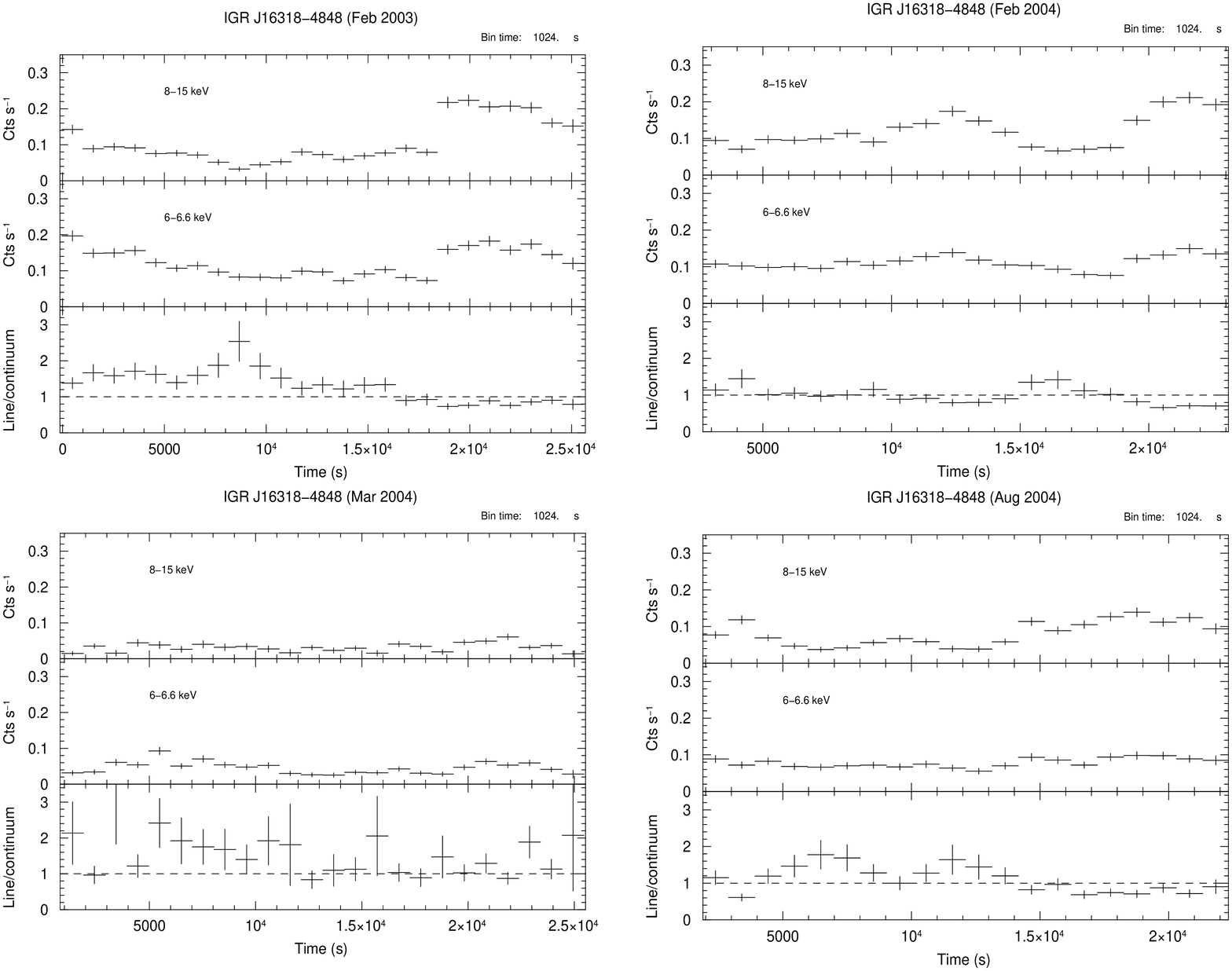}
}
\caption{XMM-Newton EPIC-pn light curves in the 8-15 keV (upper panel) and 6-6.6 keV 
(middle panel) energy ranges. In the lower panel, the ratio of the two
light curves is shown (bin time=1024 secs).}
\label{fig1}
\end{figure*}

\section{Observations and data reduction}\label{data}

\subsection{XMM-Newton}

The monitoring campaign consists of three observations spaced by
periods of one month and five months, respectively. In this paper, we
present a detailed study of these three observations along with the
reanalysis of the XMM-Newton observation performed in February 2003.
Table~\ref{tab1} summarizes the information on all XMM-Newton
observation covered by this paper, together with the EPIC-pn
total exposure time and the extraction region used for each
observation. The size of the extraction region for each observation
was selected such as to optimise the signal to noise ratio.

In this paper, we only discuss the EPIC-pn data (Str\"uder et
al. 2001) as this instrument has the largest effective area above 5
keV. We have analysed the EPIC-mos data and found that it
does not contribute significantly to the results of this paper. Also,
OM data was not considered since it is outside the scope of this
paper.

All data sets were analyzed using SAS v.6.1.0 (Gabriel et al. 2004).
Following the standard procedure for XMM-Newton data reduction, we
used the most recent calibration files available in September 2004.

The spectra were accumulated using single- and double-events 
(events where the charge in the CCD is collected in either one or two
adjacent pixels). Pile-up effects were checked using the {\it
epaplot} task and were found to be negligible in all observations. The
background contribution was always small, except in the observation of
March 2004, which was affected by high background
radiation. Time intervals for which background flaring does not
improve the signal to noise in the EPIC-pn observations were excluded
from the analysis following the method described in Piconcelli et
al. (2004). Table~\ref{tab1} shows the exposures times after
correction for background flaring activity.

All the spectra have been rebinned, in order to ensure that each
background-subtracted spectral channel has at least 25 counts, and
that the instrumental energy resolution is not oversampled by a factor
larger than 3.

\subsection{INTEGRAL}

In Table~\ref{tab2} we present the observing log of the INTEGRAL
(Winkler et al.\
2003) observations which were performed simultaneously with
XMM-Newton.  In the first two observations IGR J16318-4848 was the
main target and the observations were done using a so-called 5x5
rectangular dither pattern. In the third observation the main target
was the nearby ($\sim 56\arcmin$ away) source IGR J16320-4751
(Rodriguez et al. 2006) and the observation was done using a so-called
hexagonal dither pattern.  The 5x5 rectangular dither pattern consists
of a square pattern around the nominal target location (1 source
on-axis pointing, 24 off-source pointings, with a 2 degrees step),
while the hexagonal dither pattern consists of a hexagonal pattern
around the nominal target location (1 source on-axis pointing, 6
off-source pointings, each 2 degrees apart). IGR J16318-4848 was
always in the fully coded field of view of both the IBIS and SPI
instruments, where the instrumental response is optimal.

\begin{table}[h]
\caption{INTEGRAL monitoring campaign observations on IGR J16318-4848.
The table shows the start time and end time of IBIS/ISGRI INTEGRAL
observations on this source.}
\centering
\begin{tabular}{ccc}
\\
\hline\hline
\multicolumn{2}{c}
{\bf Start Time - End Time (UT)}     \\ \hline \hline
2004-02-18 03:41 & 2004-02-18 17:29  \\
2004-03-20 20:15 & 2004-03-21 12:35  \\
2004-08-19 13:46 & 2004-08-20 22:41  \\
\hline
\end{tabular}
\label{tab2}
\end{table}

In this paper we only consider data from IBIS/ISGRI (Lebrun et
al. 2003 and Ubertini et al. 2003) and use only those data that
overlap with the XMM-Newton observations. We do not consider the data
from the IBIS/PICsIT, SPI, Jem-X or OMC instruments, since either the
angular resolution is too high (SPI: 2.5$^\circ$) and therefore the
various sources in the Norma region of the Galaxy close to each other
make a proper analysis not possible (e.g., Rodriguez et al. 2006), or
IGR J16318-4848 was too weak to be detected (IBIS/PICsIT, Jem-X, OMC).

\begin{figure*}
\hbox{
\hspace{-0.5cm}
\includegraphics[width=14.0cm, angle=90]{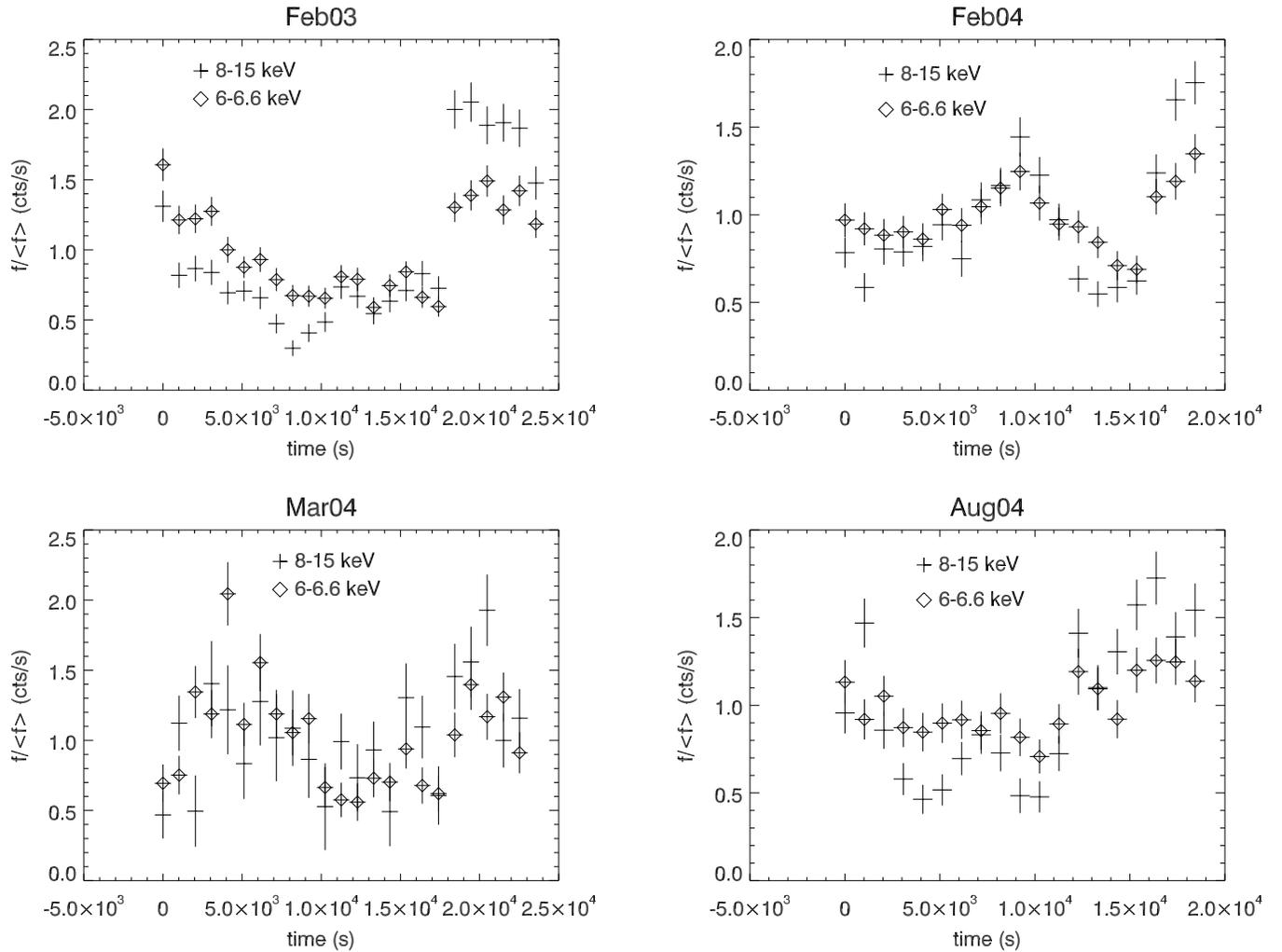}
} \caption{XMM-Newton count rate divided by the average count rate in each
band. The rapid flux increments in the February 2003, February 2004 and
August 2004 observations take places simultaneously in both bands. }
\label{fig2}
\end{figure*}

The INTEGRAL data reduction and the extraction of the events were
performed using one of the latest versions of the Off-line Scientific
Analysis ({\tt OSA}; Courvoisier et al.\ 2003), i.e., v5.0. The data
from the IBIS/ISGRI instrument were used to produce images in the
20--40 and 40--80 keV energy ranges, with the aim of effectively
identifying the most active sources in the field. From the results of
this step, we produced a catalogue of active sources which was given
as an input for a second run producing images in the 20--60 and
60--200 keV energy ranges; we increased the first range up to 60
keV to increase the signal to noise ratio in the lower energy band.
In the latter we forced the extraction/cleaning of each of the
catalogue sources in order to obtain the most reliable results for IGR
J16318-4848 (see Goldwurm et al.\ 2003 for a detailed description of
the IBIS analysis software). These results were used in the production
of the energy spectra for IGR J16318-4848.  Note that in all these
processes, in order to be completely simultaneous with the XMM-Newton,
we restricted the data reduction to the times that were strictly
simultaneous with the XMM-Newton observations (see
Table~\ref{tab1}). To do so we created good time interval fits files
that were provided as an input to OSA; we also restricted the spectral
energy range to 20--60 keV, since no significant emission was
detected above these energies.

\section{Results}\label{results}
\subsection{Timing analysis}

In Fig.~\ref{fig1}, we show the IGR~J16318-4848 light curves in the
energy ranges, 8-15 keV (upper panel) and 6-6.6 keV (middle panel),
together with the ratio of the two light curves (lower panel). We
used a bin time of 1024 secs, although we also studied several other
bin times, such as, 256 and 512 secs, but they do not provide further
improvement on the data quality. In this plot we can clearly see the
February 2003 high energy flare that was studied by Matt \& Guainazzi
(2003) and variations by a factor $2\sim3$ in $\leq 5 \times 10^{3}$ s
are visible also in the February 2004 light curve, and a short flare
in the August 2004 light curve.

Only the February 2003 observation shows a significant variation in
the ratio of the two light curves. This effect could indicate
variations of the properties of the cold matter on times scales of
about $10^{4}$ s (Matt \& Guainazzi, 2003). This variation in the
continuum flux with respect to the flux in the Fe K$_\alpha$ line is
not present in the rest of the observations.

\begin{figure}
\hbox{
\hspace{-0.1cm}
\includegraphics[width=6.4cm, angle=90]{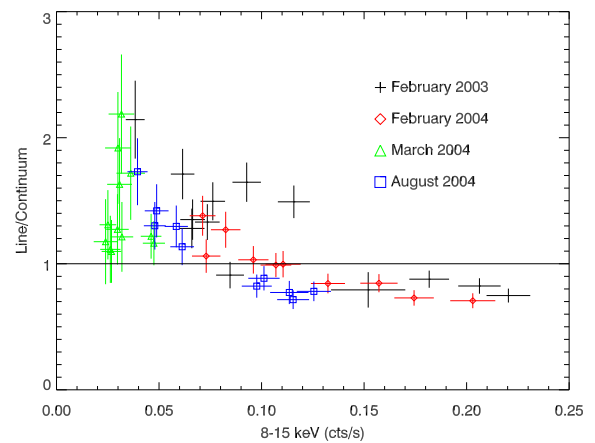}
} \caption{Ratio between the 6-6.6 keV and the 8-15 keV energy band as
a function of the 8-15 keV count rate.}
\label{fig3}
\end{figure}

In Fig.~\ref{fig2}, we show the count rate in the 8-15 keV energy
range, which is dominated by the continuum, and the count rate in the
6-6.6 keV energy range, which is dominated by the Fe K$_\alpha$ line,
divided by the average count rate in each band.  During the February
2003, February 2004 and August 2004 observations, a flux increment can
be clearly seen. In all cases the Fe K$_{\alpha}$ line follows almost
immediately the continuum. This suggests that the emission lines are
produced by the illumination of optically-thick matter on a small
scale around the high-energy continuum source. These results are
consistent with Walter et al. (2003). However, the variation of the
continuum exhibits a larger dynamical range than the variation of the
Fe K$_{\alpha}$ iron line (see Fig.~\ref{fig3}). This suggests that at
least part of the line emission comes from regions, which do not
respond on small timescales to the continuum.

\begin{table*}
\caption{Best-fit parameters and results of the IGR~J16318-4848 monitoring campaign using the {\it  baseline model}. }
\begin{center}
\begin{tabular}{lcccc} \hline \hline
Parameter & February 2003 & February 2004 & March 2004 & August 2004 \\ \hline
$\Gamma$                                & $1.54 ^{+0.02}_{-0.4}$ & $1.46 \pm 0.03$ & $1.35 \pm 0.04 $ & $1.41 \pm 0.03$ \\
$E_c$ [keV]                             & $58 ^{+30}_{-11}$ & $<16$ & $<180$ & $<30$ \\
A [$10^{-2}$~cm$^{-2}$~s$^{-1}$]        & $35 \pm 1 $ & $27.0 ^{+0.8}_{-0.7} $ & $9.95 \pm 0.01 $ & $4.8 ^{+0.1}_{-0.2}$ \\
HR[$\frac{20-30}{30-50}$]~[keV]   & ... & $0.40 \pm 0.14$ & $0.46 \pm 0.04$ & $0.44 \pm 0.10$ \\
$N_H$~[$10^{24}$~cm$^{-2}$]             & $2.12 \pm 0.03 $ & $1.82 ^{+0.05}_{-0.03} $ & $2.16 ^{+0.07}_{-0.06} $ & $1.38 ^{+0.04}_{-0.03}$ \\
$E_1$~(Fe K$_{\alpha1}$)~[keV]           & $6.39 \pm 0.01$ & $6.43 ^{+0.01}_{-0.02} $ & $6.44 \pm 0.01 $ & $6.42 ^{+0.05}_{-0.04} $ \\
$I_1$~(Fe K$_{\alpha}$)~[$10^{-4}$~cm$^{-2}$~s$^{-1}$] & $1.71 \pm 0.02 $ & $0.93 \pm 0.04 $ & $0.83 ^{+0.04}_{-0.02}$ & $0.65 ^{+0.06}_{-0.03}$ \\
EW (Fe K$_{\alpha}$) [eV]               & $9.5 \pm 0.3 $ & $8.0 \pm 0.4 $ & $8.6 ^{+0.6}_{-0.5}$ & $23.9 ^{+2.1}_{-1.1}$ \\
$E_2$~(Fe K$_{\beta}$)~[keV]            & $7.05 \pm 0.01 $ & $7.10 ^{+0.03}_{-0.02}$ & $7.05 ^{+0.02}_{-0.01}$ & $7.05 ^{+0.04}_{-0.02}$ \\
$I_2$$^a$                               & $0.20 \pm 0.02$ & $0.24 ^{+0.03}_{-0.06} $ & $0.18 ^{+0.03}_{-0.05}$ & $0.25 \pm 0.05 $ \\
EW (Fe K$_{\beta}$) [eV]                & $2.2 ^{+0.3}_{-0.2} $ & $2.3 ^{+0.3}_{-0.5} $ & $2.8 ^{+0.3}_{-0.7}$ & $7.1 \pm 1.4 $ \\
$E_3$(Ni K$_{\alpha}$)~[keV]            & $7.45 ^{+0.06}_{-0.01}$ & $7.45 ^{+0.1}_{-0.01} $ & $7.5 ^{+0.4}_{-0.1}$ & $7.5 ^{+0.4}_{-0.7}$ \\
$I_3$$^a$                               & $0.05 ^{+0.02}_{-0.01} $ & $0.05 ^{+0.03}_{-0.02} $ & $0.03 ^{+0.02}_{-0.01} $ & $0.07 ^{+0.03}_{-0.05} $ \\
EW (Ni K$_{\alpha}$) [eV]               & $0.7 ^{+0.2}_{-0.1} $ & $0.5 ^{+0.4}_{-0.2} $ & $0.5 ^{+0.4}_{-0.3}$ & $2.2 ^{+0.8}_{-1.3}$ \\
$I_{CS}$$^a$                            & $0.05 \pm 0.03 $ & $0.25 \pm 0.06 $ & $0.24 ^{+0.13}_{-0.06} $ & $0.19 ^{+0.08}_{-0.07}$ \\
$\chi^2$/$\nu$                          & 106.9/92 & 98.9/97 & 64.9/51 & 111.7/83 \\
$\chi^2$/$\nu$ (Without~CS) 		& $120.9/93$ & $153.0/98$ & $114.85/52$ & $135.85/84$ \\
$F$~[10$^{-11}$~erg~cm$^{-2}$~s$^{-1}$] & $2.39 ^{+0.03}_{-0.06}$ & $ 2.11 \pm 0.08 $ & $0.8 ^{+0.5}_{-0.2}$ & $1.2 ^{+0.3}_{-0.2}$ \\
$CR~(1-5 keV)~[10^{-3}~$cts$~$s$^{-1}]$ & $ 1.3 \pm 0.6 $ & $ 12.9 \pm 1.4 $ & $ < 2.5 $ & $ 2.10 \pm 0.40 $ \\
\hline \hline
\end{tabular}

\noindent
$^a$ratio intensity against the Fe K$_{\alpha}$\\

\label{tab3}
\end{center}
\end{table*}

\subsection{Spectral analysis}

\begin{figure}
\hbox{
\hspace{-0.1cm}
\includegraphics[width=6.4cm, angle=-90]{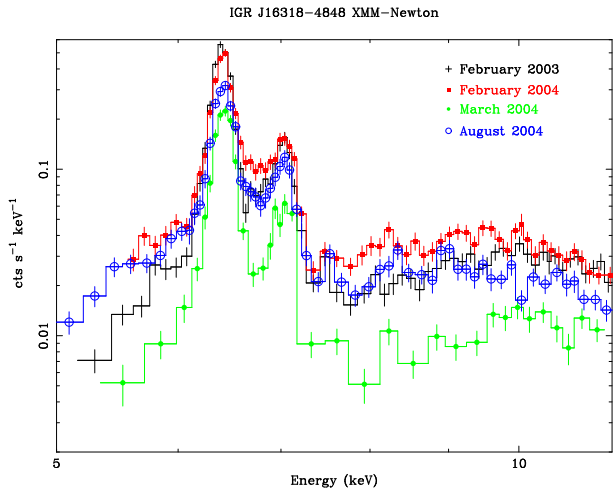}
} \caption{Spectra of the four IGR~J16318-4848 XMM-Newton
observations. Despite the large flux variations between different
observations, the spectral shapes follow the same trend.}
\label{fig4}
\end{figure}

In Fig.~\ref{fig4} all four IGR~J16318-4848 spectra are shown. Despite
the large flux variability, the spectral shapes are very similar one
another. All the spectral analysis has been made using events above 5
keV because the source is highly absorbed below this energy.
Detections at a level larger than 3 $\sigma$ have been found only in
the February 2004 and August 2004 observations (in Table~\ref{tab3} we
shown the count rate (CR) in the 1-5 keV energy band).

\begin{figure}[hbt]
\hbox{
\hspace{-0.1cm}
\includegraphics[width=7.4cm, angle=+90]{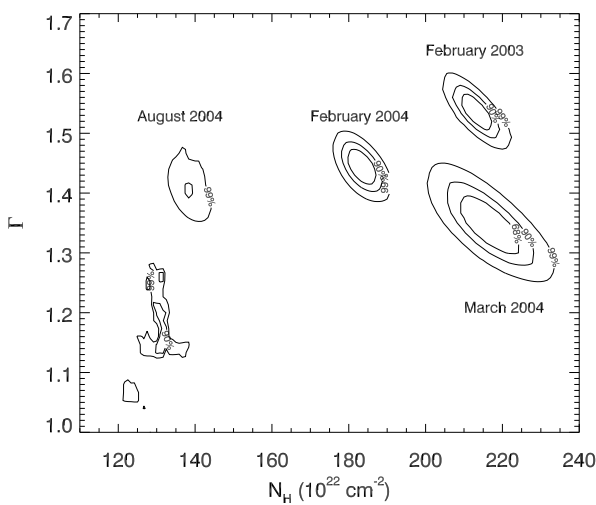}
} \caption{$\Gamma-N_{H}$ ISO-$\chi^2$ contour plots corresponding to IGR
J16318-4848 XMM-Newton observations.}
\label{fig5}
\end{figure}

Following Matt \& Guainazzi (2003), the spectra of all available
XMM-Newton observations of IGR~J16318-4848 have been fitted (using
XSPEC v.11.3.1), alongside the simultaneous INTEGRAL spectra
whenever available, with the following {\it baseline model}:
$$
F(E) = e^{-N_H \sigma(E)} [AE^{-\Gamma} e^{E/E_c} + \sum_{i=1}^4 G_i(E) + \sum_{i=1}^2 S_{cs_i}(E)]
$$

\begin{figure*}
\hbox{
\hspace{-0.1cm}
\includegraphics[width=1.0\textwidth, angle=0]{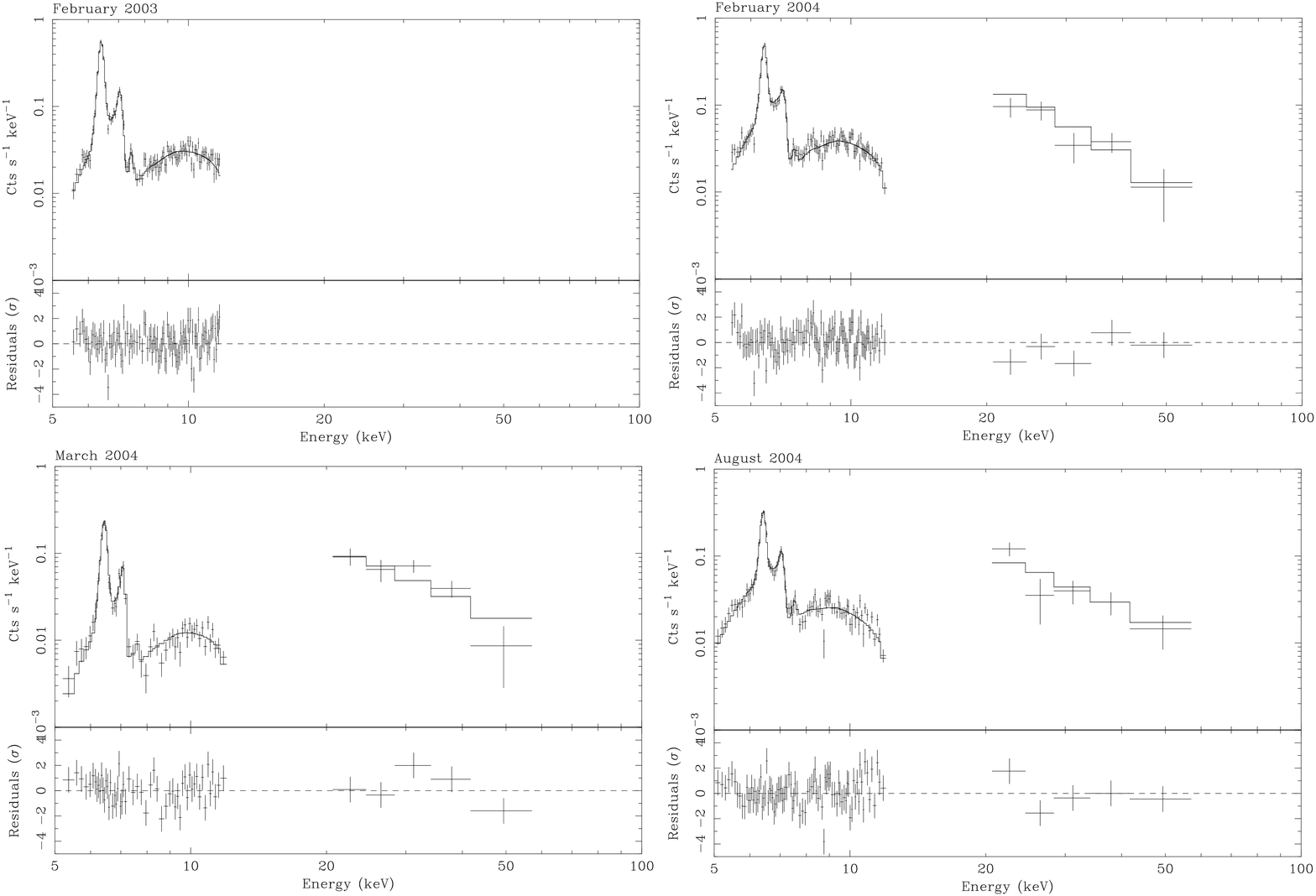}
}
\caption{Spectra ({\it upper panels}) and
residuals in units of standard deviations
({\it lower panels}) when the baseline model
is applied to the IGR~J16138-4848 observations
discussed in this paper.}
\label{fig6}
\end{figure*}

\noindent where $\sigma(E)$ is the sum of the photoelectric and the
Thompson cross-sections, $G_i(E)$ are Gaussian profiles corresponding
to the Fe K$_{\alpha}$ doublet, Fe K$_{\beta}$ and Ni K$_{\alpha}$
fluorescent emission lines, and $S_{cs_i} (E)$ is the Fe K$_{\alpha
  1}$ and Fe K$_{\alpha 2}$ Compton Shoulder respectively. The energy first
component of the Fe K$_{\alpha}$ doublet has been left free to vary in
the fit, while the energy centroid of the second component of the Iron
line doublet was fixed to the energy of the first component multiplied
by 1.002 (Kranse et al. 1979), as for neutral iron. The relative intensity of the second
component of the Iron doublet was fixed to twice the intensity of the
first component, as dictated by atomic physics.


The last component of the {\it baseline model} was modeled using two
box functions  (as a first order approximation) corresponding to
the Compton Shoulder (CS) of Fe K$_{\alpha}$ doublet 
(e.g. Sunyaev \& Churazov 1996, Matt 2002; Watanabe et al. 2003). The
first box function corresponding to the lowest energy component of the
iron doublet was extended in the closed interval
[$E_1$-0.16~keV,$E_1$-0.01~keV] as appropriate for the one-scattering
CS (Matt 2002), where $E_1$ is the best-fit centroid energy for the Fe
K$_{\alpha 1}$ line,  and 0.16 keV is the maximum recoil energy.
We neglected the higher order shoulders, i.e. the ones dominated by
two or more scatterings, because they are much fainter even for
Compton--thick media, both in the reflection (Matt et al. 1991; George
\& Fabian 1991) and in the transmission (Leahy \& Creighton 1993)
cases. Similarly, the box model for the second component was extended
in the closed interval [$E_2$-0.16~keV,$E_2$-0.01~keV].

In principle, one should include in the fit the Compton Shoulder for
the other fluorescent lines as well.  However, these lines are much
fainter, and their Compton Shoulder are basically undetectable in the
EPIC-pn spectra. Therefore, for the sake of simplicity 
they will not be considered in what follows. For the same reason, the 
Fe K$_{\beta}$ and Ni K$_{\alpha}$ lines were modeled as single Gaussians.

Regarding the absorbing matter, the abundance (normalized to the solar
value) of elements with $Z \ge 26$, ($Z_{\rm Fe}$ hereafter) was
allowed to vary independently of the abundance of the lighter
elements. The relative iron abundance values (Anders E. \& Grevesse
N.) obtained for each observation were: $0.81\pm0.03$,
$0.77\pm^{0.03}_{0.04}$, $0.89\pm^{0.05}_{0.06}$ and
$0.86\pm^{0.05}_{0.06}$ for the February 2003, February 2004, March
2004 and August 2004 observations, respectively. These values are
compatible within the errors and therefore we calculated the weighted
mean, which resulted to be $0.81\pm0.02$, and used it in all following
spectral fits.

XMM-Newton and INTEGRAL spectra were fitted simultaneously applying
the relative normalization between the EPIC-pn and ISGRI data
given by Kirsch et al. (2005).

We studied the width of the emission lines in order to avoid any
possible numerical problem related to unresolved lines. Firstly, we
fit all the spectra with the line width fixed to $10^{-4}$ eV. We then
allowed the line width of the emission lines to vary, but no
improvements in the $\chi^2$ was found. With these conditions, we
obtained very tight upper limits for the width of the emission lines,
smaller than 10 eV. The only exception is the August 2004 observation,
when the width is basically unconstrained.  We then fixed the line
width to $10^{-4}$ eV for all the observations.

Despite the strong coupling between $\Gamma$ and $N_{H}$ that was
found in the February 2003 observations (Matt \& Guainazzi, 2003),
we have been able to find a stable spectral index for all the
observations, with the exception of August 2004 (see Fig.~\ref{fig5}). 
This has been posible due to the large energy coverage introduced 
by the simultaneous INTEGRAL data; the better constraint on the metal
abundance and the inclusion in our model of the Fe K$_{\alpha}$ doublet
(see Fig.~\ref{fig5}).

In Table~\ref{tab3} we summarize the best-fit parameters of the {\it
baseline model} and in Fig.~\ref{fig6} we present the spectra and
residuals of the data against the {\it baseline model}. All the errors
shown in this paper correspond to a 90\% confidence level for one
interesting parameter. The values of the equivalent widths in the
table were computed with respect to the unabsorbed continuum.

In all 2004 observations, the relative intensity of the Compton
Shoulder is larger than in the 2003 observation. Interestingly, for
these observations the centroid of the Fe K$_{\alpha1}$ line is
significantly bluer than expected for neutral iron (6.40~keV). This
seems to suggest significant ionization of iron (Fe$~\geq$ {\sc XIV},
Kallman et al. 2004).  However, the strong iron Fe K$_{\beta}$ line
implies that iron is less ionized than Fe {\sc xi} or so (Molendi et
al, 2003). The apparent blueshift may then be an instrumental effect,
or an artifact due to either the oversimplified CS modeling or to the
assumption of a narrow line core. We recall that the systematic
uncertainties on the gain of the EPIC-pn camera at 6 keV is 10 eV (Kirsch 2006).

\begin{figure}
\hbox{
\hspace{-0.1cm}
\includegraphics[width=6.4cm, angle=-90]{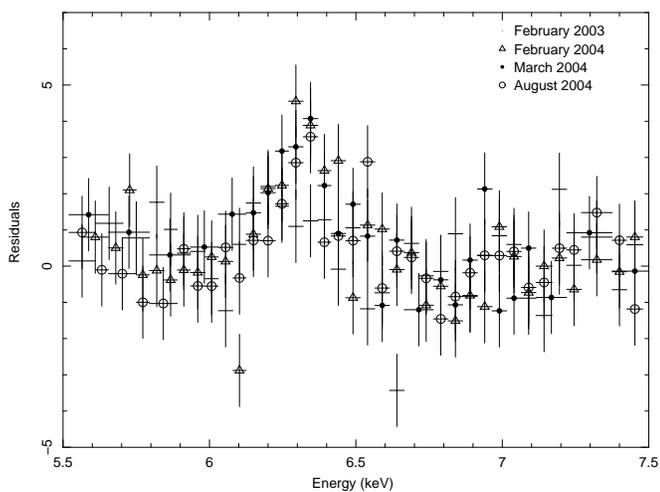}
} \caption{Residuals of the Compton--Shoulder feature for all of the
IGR J16318-4848 observations.}
\label{fig7}
\end{figure}

In Fig.~\ref{fig7} we show the residuals of the Compton--Shoulder
feature for all the observations. We have also included in the
Table~\ref{tab3} the $\chi^2$ values of our {\it baseline model}
without the Compton Shoulder. In all the observations the improvement
adding the CS is statistically significant.

A comparison between the ratio of all INTEGRAL spectra and their mean
spectrum was performed.  showing no trend. This stable spectral shape
agrees with the accreting pulsar objects.

Interestingly, Matt \& Guainazzi (2003) report soft excess emission
above the extrapolation of the baseline model at energies $E \approxlt
5$~keV. We have reanalyzed the images obtained during the monitoring
campaign and discovered a serendipitous source at $\simeq$30$\arcsec$ from
the IGR~J16318-4848 centroid position. This serendipitous source is
totally responsible for the reported soft excess. In the other three
observations, this soft source falls into the CCD gaps. Still, we
detect soft X-ray emission at a significance level larger than 3
$\sigma$ in two out of the four observations. No spectral information
can be derived on this component from the available data.

\section{Discussion}\label{discussion}

We have presented results from a coordinated XMM-$Newton$/INTEGRAL
monitoring campaign on IGR~J16318-4848. The source was always clearly
detected, albeit with large amplitude (factor of 3) flux variations,
suggesting (along with its rediscovery in the ASCA archive) that it is
a persistent source which escaped attention before 2003 only due to
the strong absorption. Persistence of the X-ray emission is a
characteristic of Be systems with wide orbits ( hundreds of days,
Negueruela 1998).

\begin{figure}[hbt]
\hbox{
\hspace{-0.1cm}
\includegraphics[width=9.0cm, angle=0]{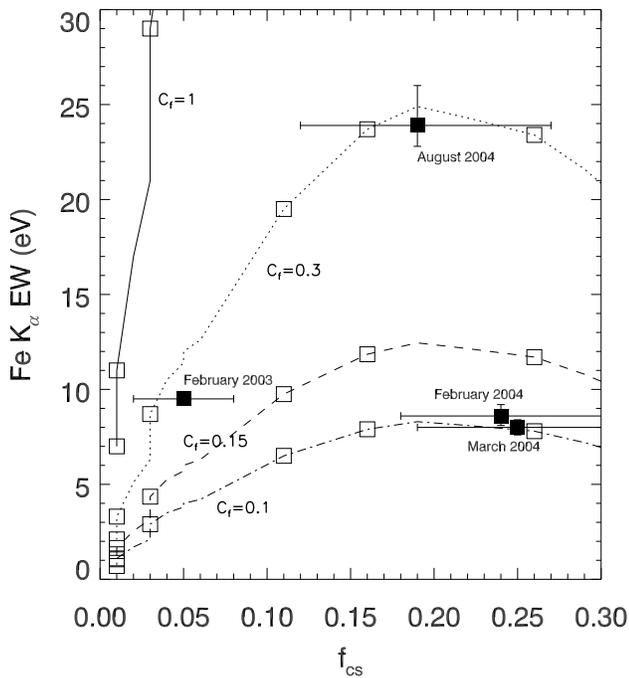}
} \caption{Fe K$_{\alpha}$ fluorescent
line EW against the Compton  Shoulder normalized intensity ($f_{CS}$). The {\it
lines} represent the predictions of radiative transfer simulations after Matt
(2002) for different values of the gas covering fraction, $C_f$. The {\it filled
squares} represent the four XMM-Newton measurements discussed in this paper.
The {\it empty squares} indicate
the position along the lines corresponding to the following values of column
density of the obscuring gas: $N_H = 10^{22}$, $2 \times 10^{22}$, $5 \times
10^{22}$, $10^{23}$, $2 \times 10^{23}$, and $5 \times 10^{23}$~cm$^{-2}$.
}
\label{fig8}
\end{figure}

The properties of the circumstellar matter clearly changed from one
observation to the other, the column density (as derived from absorption)
for instance being 
significantly lower (but still Compton-thick) in the last observation.
As a further probe of insight on the circumstellar matter, we used as a 
diagnostic the Compton Shoulder, which depends on the covering fraction 
of the matter and on its {\sl average} column density (Matt 2002, Matt \&
Guainazzi 2003), while the absorption of course gives information on
 the line-of-sight
(l.o.s.) column only. 

In Fig.~\ref{fig8} we show the results of the Fe K$_\alpha$ Compton Shoulder
normalized intensity, $f_{CS}$ (the ratio between the CS and the 
line core intensities) against
the EW of the Fe K$_\alpha$. Comparing our results with the predictions of 
radiative transfer simulations done in Matt (2002), a covering
fraction of the obscuring matter in the range 0.1-0.3, and average column
densities in the range $2 \times 10^{22}$ to $5 \times 10^{23}$~cm$^{-2}$ 
are derived. 

It is worth noting that the average column density is always much
smaller (1 to 2 orders of magnitude) than the l.o.s. one. The fact
that the two values are largely different implies that the
circumstellar matter is highly inhomogeneous. As the average one is
systematically smaller that of the l.o.s. it strongly suggests that we
are observing the source from a privileged direction. This, together
with the relatively small covering factor (which rules out the
hypothesis that the compact object  (a neutron star according to
Filliatre \& Chaty 2004, c.f., Negeruela 1998) is embedded in the
wind from the giant companion  (a sgB[e] star according to
Filliatre \& Chaty 2004), not surprisingly if the system has indeed a
wide orbit, as suggested by its persistency), can be interpreted in
terms of a flattened distribution of matter (quite plausible in a
binary system, especially if with wide orbit), with the column
density decreasing from the equatorial plane, and the line of sight
lying close to that plane. Even if this scenario is by no means
unique, it seems to be the simplest and most natural. The variations
in the covering factor and average column density may then be the
witness of variations in the mass transfer rate, possibly due to a
highly eccentric orbit of the system.

\begin{acknowledgements}
This paper is based on observations obtained with XMM-Newton, an ESA
science mission with instruments and contributions directly funded by
ESA Member States and the USA (NASA).

\end{acknowledgements}

\end{document}